\documentclass[preprint,showpacs,titlepage,aps,prd,
amsmath,byrevtex,nofootinbib]{revtex4}

\usepackage{graphicx,amssymb}

\def\lsim{\raise0.3ex\hbox{$\;<$\kern-0.75em\raise-1.1ex
\hbox{$\sim\;$}}}
\def\gsim{\raise0.3ex\hbox{$\;>$\kern-0.75em\raise-1.1ex
\hbox{$\sim\;$}}}

\DeclareMathAlphabet{\mathsc}{OT1}{cmr}{m}{sc}

\begin{document}
\voffset=-0.5cm

\baselineskip 0.58cm

\preprint{ Fermilab-Pub-05-049-T}
\preprint{hep-ph/0601198}

\vglue -0.1cm
\title{\Large What Fraction of Boron-8 Solar Neutrinos
arrive\\ at the Earth as a $\nu_2$ mass eigenstate?\footnote{Dedicated to the memory of John Bahcall who championed solar neutrinos for many lonely years.} }

\author{Hiroshi Nunokawa$^1$} \email{nunokawa@fis.puc-rio.br}
\author{Stephen Parke$^2$} \email{parke@fnal.gov} 
\author{Renata  Zukanovich Funchal$^3$} \email{zukanov@if.usp.br} \affiliation{
  $^1$\sl Departamento de F\'{\i}sica, Pontif{\'\i}cia Universidade Cat{\'o}lica do Rio de Janeiro, \\
  C. P. 38071, 22452-970, Rio de Janeiro, Brazil \\
  $^2$Theoretical Physics Department, Fermi National Accelerator Laboratory,\\
  P.O. Box 500, Batavia, IL 60510, USA\\
  $^3$Instituto de F\'{\i}sica, Universidade de S\~ao Paulo, C.\ P.\
  66.318, 05315-970 S\~ao Paulo, Brazil}

\date{January 24, 2006}

\begin{abstract}
  We calculate the fraction of $^8$B solar neutrinos that arrive at
  the Earth as a $\nu_2$ mass eigenstate as a function of the neutrino energy.
  Weighting this fraction with the $^8$B neutrino energy
  spectrum and the energy dependence of the cross section for the
  charged current interaction on deuteron with a threshold on the kinetic energy of the recoil
  electrons of 5.5 MeV,
 we find that the integrated weighted fraction of $\nu_2$'s to be
  91$\pm$2\% at the 95\% CL. 
  This energy weighting procedure corresponds to the charged current response 
  of the Sudbury Neutrino Observatory (SNO).  We have used 
SNO's current best fit values for the solar mass squared difference 
  and the mixing angle, obtained by combining the data from all solar 
  neutrino experiments and the reactor data from KamLAND.
The   uncertainty on the $\nu_2$ fraction
  comes primarily from the uncertainty on the solar $\delta m^2$
  rather than from the uncertainty on the solar mixing angle or the Standard Solar Model. 
  Similar results for the Super-Kamiokande experiment are also given.
  We extend this analysis to three neutrinos and discuss how to extract
  the modulus of the Maki-Nakagawa-Sakata mixing matrix element U$_{e2}$
  as well as place a lower bound
on the electron number density in the solar $^8$B neutrino production region. 
\end{abstract}

\maketitle

\baselineskip 0.75cm

\section{Introduction}

Recently the KamLAND~\cite{KamLAND} and Sudbury Neutrino Observatory (SNO)~\cite{SNO} experiments
have given a precise determination of the neutrino solar mass squared
difference and mixing angle responsible for the solar neutrino deficit
first observed in the Davis~\cite{Davis} experiment when compared to
the theoretical calculations by Bahcall~\cite{Bahcall}.  
Subsequently this deficit has been observed by many other 
experiments~\cite{SK_solar, solar_other}, while the theoretical 
calculations of the neutrino flux based on the Standard Solar Model (SSM) 
has been significantly improved\cite{SSMs}. 
When all of these results
are combined in a two neutrino fit as reported by SNO~\cite{SNO},
the allowed values for the solar mass squared difference, $\delta
m^2_{\odot}$, and the mixing angle, $\theta_{\odot}$, are individually
(for 1 degree of freedom) restricted to the following
range\footnote{We use the notation of ~\cite{sinsq} with the subscript
  ``$\odot$'' reserved for the two neutrino analysis whereas the
  subscript ``$12$'' is reserved for the three neutrino analysis.},
\begin{eqnarray}
 \delta m^2_{\odot} & = & 8.0^{+0.4}_{-0.3} \times 10^{-5} {\rm eV^2},
 \nonumber \\ \sin^2 \theta_{\odot} & = & 0.310 \pm 0.026,
\label{solar}
\end{eqnarray}
at the 68 \% confidence level.  Maximal mixing, $\sin^2 \theta_\odot
=0.5 $, has been ruled out at greater than 5 $\sigma$.  The solar
neutrino data is consistent with $\nu_e \rightarrow \nu_\mu ~{\rm
  and/or} ~\nu_\tau$ conversion.
The precision on $ \delta m^2_{\odot}$ comes primarily from the
KamLAND experiment~\cite{KamLAND} whereas the precision on
$\sin^2\theta_{\odot}$ comes primarily from the SNO
experiment~\cite{SNO}.

The physics responsible for the reduction in the solar $^8$B electron
neutrino flux is the Wolfenstein matter effect~\cite{Wolf} with the
electron neutrinos produced above the Mikeyev-Smirnov (MS)
resonance~\cite{MS}.  The combination of these two effects in the
large mixing angle (LMA) region, given by Eq.~(\ref{solar}), implies
that the $^8$B solar neutrinos are produced and  propagate adiabatically to
the solar surface, and hence to the earth, as almost a pure $\nu_2$
mass eigenstate.\footnote{Without the matter effect, the fraction of $\nu_2$'s
would be simply $\sin^2 \theta_{\odot}$, i.e. about 31\%, and energy independent.}
Since, approximately one third of the 
$\nu_2$ mass eigenstate is
$\nu_e$, this explains the solar neutrino deficit first
reported by Davis.  If the $^8$B solar neutrinos arriving at the Earth
were 100\% $\nu_2$, then the {\em day-time} Charged Current (CC) to Neutral
Current (NC) ratio, CC/NC, measured by SNO would be exactly $\sin^2
\theta_\odot$, the fraction of $\nu_e$ in $\nu_2$ in the two neutrino
analysis.

Of course, the $\nu_2$ mass eigenstate purity of the solar $^8$B
neutrinos is not 100\%, as we will see later, some fraction arrive as
$\nu_1$'s and if the electron neutrino has a non-zero component in
$\nu_3$ (i.e. non-zero $\sin^2 \theta_{13}$) then there will be a
small fraction arriving as $\nu_3$'s. 
For all practical solar neutrino experiments, these mass eigenstates can be considered
to be incoherent, see \cite{DLS}.
The mass eigenstate purity of
the $^8$B solar neutrinos is the main subject of this paper.  In the
next section we will summarize the important physics of the MSW-LMA
solar neutrino solution outlined above and calculate the mass
eigenstate purity of $^8$B neutrinos as a function of the neutrino
energy in a two neutrino analysis for both the SNO and Super-Kamiokande (SK) experiments.  
In section 3 we will discuss what
happens in a full three neutrino analysis.  In section 4, as an
application of the previous sections, we will discuss the possibility
of extracting information about the solar interior independently
from the standard solar model.  Finally, in section 5, we present our
summary and conclusions.

\section{Two Neutrino Analysis:}

\subsection{$^8$B $\nu_2$ Fraction}

\begin{figure}[t]
\centering\leavevmode
\vglue 0.2cm
\includegraphics[width=11.cm]{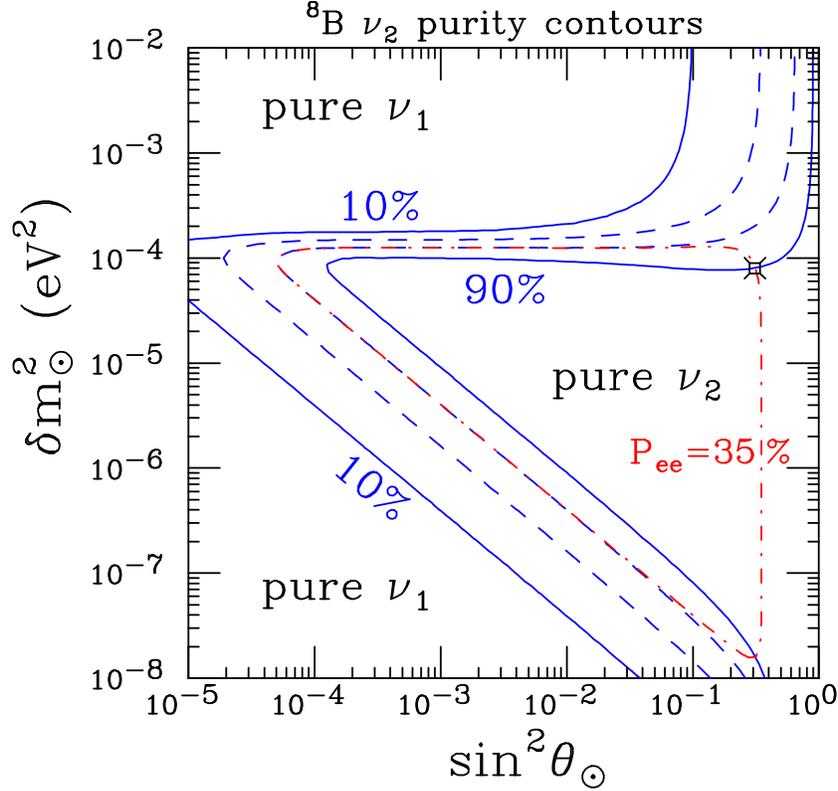}
\vglue -0.5cm
\caption{The solid and dashed (blue) lines are the 90, 65, 35 and 10\% 
iso-contours of the fraction of the solar $^8$B neutrinos that are $\nu_2$'s
  in the $\delta m^2_\odot$ and $\sin^2 \theta_\odot$ plane.  
  The current best fit value, indicated by the
  open circle with the cross, is close to the 90\% contour. The iso-contour for an electron 
  neutrino survival probability, $P_{ee}$, equal to 35\% is the dot-dashed (red)
  ``triangle'' formed by the 65\% $\nu_2$ purity contour for small $\sin^2 \theta_\odot$ 
  and a vertical line in the pure $\nu_2$ region at
  $\sin^2 \theta_\odot=0.35$. 
  Except at the top and bottom right hand corners of this triangle 
  the  $\nu_2$ purity is either 65\% or 100\%.
  }
\label{nu1nu2}
\end{figure}

In the two neutrino analysis, let $f_1(E_\nu)$ and $f_2(E_\nu)$ be the
fraction of $^8$B solar neutrinos of energy $E_\nu$ which exit the Sun
and thus arrive at the Earth's surface as either a $\nu_1$ or a
$\nu_2$ mass eigenstate, respectively.  Following the analytical
studies of Ref.~\cite{Parke86}, these fractions are given by
\begin{eqnarray}
f_1(E_\nu) & = & \langle \cos^2 \theta_\odot^N 
- P_x \cos 2\theta_\odot^N \rangle_{^8\text{B}}, \\
f_2(E_\nu) & = & \langle  \sin^2 \theta_\odot^N 
+ P_x\cos 2 \theta_\odot^N \rangle_{^8\text{B}},
\end{eqnarray}
where $\theta_\odot^N$ is the mixing angle defined at the $\nu_e$
production point, $P_x$ is the probability of the neutrino to jump
from one mass eigenstate to the other during the MS-resonance
crossing, and the sum is constrained to be 1, $f_1+f_2=1$.  
The average $\langle \cdots \rangle_{^8\text{B}}$ 
is over the electron density of the
$^8$B $\nu_e$ production region in the center of the Sun
predicted by the Standard Solar Model~\cite{BS2005}. 
The mixing angle, $\theta_\odot^N$, and the mass difference squared, $\delta m^2_N$,
at the production point are
\begin{eqnarray}
 \sin^2 \theta_\odot^N & = & \frac{1}{2} \left\{ 1 + \frac{(A-\delta m^2_\odot  \cos 2\theta_\odot)}
 {\sqrt{(\delta m^2_\odot \cos 2\theta_\odot - A)^2 + (\delta m^2_\odot \sin 2 \theta_\odot)^2}} \right\},
 \label{sinsqthN} \\[0.5cm]
 \delta m^2_N & = & \sqrt{(\delta m^2_\odot  \cos 2\theta_\odot-A)^2+(\delta m^2_\odot \sin 2 \theta_\odot)^2}
\end{eqnarray}
where
\begin{equation}
 A  \equiv 2\sqrt{2}G_F (Y_e \rho/M_n) E_\nu = 1.53\times 10^{-4}
{\rm eV^2} \left(\frac{Y_e \rho ~E_\nu}{ \text{kg.cm}^{-3} \text{MeV}}\right),
 \label{Adefn}
\end{equation}
is the matter potential, $E_\nu$ is the neutrino energy, $G_F$ is the Fermi constant,
$Y_e$ is the electron fraction (the number of electron per nucleon),
$M_n$ is the nucleon mass and $\rho$ is the matter density. The combination
$Y_e \rho/M_n$ is just the number density of electrons.

Fig.~\ref{nu1nu2} shows, for a wide range of $\delta m^2_\odot$ and
$\sin^2 \theta_\odot$, the iso-contours of
\begin{equation}
f_2 \equiv \langle f_2(E_\nu) \rangle_E , 
\end{equation}
where $\langle \cdots \rangle_E$ is the average over the $^8$B
neutrino energy spectrum~\cite{Ortiz2000} convoluted with the energy
dependence of the CC interaction $\nu_e + d \to p + p +
e^-$ cross section~\cite{x-sec} at SNO with the threshold on the recoil electron's kinetic energy of 
5.5 MeV.
Here we use $\sin^2\theta_\odot$ as the metric for the mixing angle as
it is the fraction of $\nu_e$'s in the vacuum $\nu_2$ mass eigenstate.
In this work, we mainly focus on SNO rather than SK 
since the former is the unique solar neutrino experiment which can measure 
the total active $^8$B neutrino flux as well as $^8$B electron neutrino flux, 
independently from the SSM prediction and other experiments. 
However, we give a brief discussion on SK later in this section.

In the LMA region the propagation of the neutrino inside the Sun is
highly adiabatic~\cite{MS,Parke86,pxzero}, i.e. $P_x \approx 0$,
therefore,
\begin{eqnarray}
f_2(E_\nu)   \equiv 1-f_1(E_\nu) &=& \langle \sin ^2 \theta_\odot^N
\rangle_{^8\rm B}.
\end{eqnarray} 
Due to the fact that $^8$B neutrinos are produced in a region where
the density is significantly higher (about a factor of four)
than that of the MS-resonance value, the average $\langle
f_2(E_\nu) \rangle_E$ is close to 90\% for the current solar best fit
values of the mixing parameters from the recent KamLAND plus SNO analysis~\cite{SNO}.  
Since $\sin ^2 \theta_\odot^N \to 1$ when
$A/\delta m^2_\odot \to \infty$ (see Eq.~(\ref{sinsqthN})), we can see
that at the high energy end of the $^8$B neutrinos $ \langle \sin ^2
\theta_\odot^N \rangle_{^8 \rm B}$ must be close to 1.

We can check our result using the analysis of SNO with a simple back
of the envelope calculation.  In terms of the fraction of $\nu_1$ and
$\nu_2$ the {\it day-time} CC/NC of SNO, which is equal to 
the day-time average $\nu_e$ survival probability, $\langle P_{ee} \rangle$, 
is given by
\begin{eqnarray} 
\left. \frac{\text{CC}}{\text{NC}}\right|_{\text{day}} &=&  
\langle P_{ee} \rangle = 
f_1 \cos^2\theta_\odot + f_2 \sin^2 \theta_\odot   
\label{cctonc}, 
\label{f1eqn}
\end{eqnarray}
where $f_1$ and $f_2$ are understood to be the $\nu_1$ and $\nu_2$ 
fractions, respectively, averaged over the $^8$B neutrino energy 
weighted with the CC cross section, as mentioned before. 
Using the central values reported by SNO \footnote{For 
the sake of simplicity and transparentness of 
the discussion, we have avoided the Earth matter effect which causes 
the so called regeneration of $\nu_e$ during night, 
by simply restricting our analysis to the day time neutrino flux
throughout this paper. 
We note that due to the large error, the observed night-day asymmetry 
at SNO is consistent with any value from -8 to 5\%~\cite{SNO}
whereas the expected night-day asymmetry, 2(N-D)/(N+D), is about 2.2-3.5\% for 
the current allowed solar mixing parameters~\cite{Fogli05}. Thus the difference between
the day and the day plus night average CC/NC is less than 2\% and much smaller than SNO's 
10\% measurement uncertainty on CC/NC.
},
\begin{equation}
\left.
\frac{\text{CC}}{\text{NC}}\right|_{\text{day}} 
= 0.347 \pm 0.038,
\end{equation}
which was obtained from Table XXVI of Ref.~\cite{SNO}, and the current
best fit value of the mixing angle, we find $f_2=(1-f_1) \approx
90\%$, as expected. 
Due to the correlations in the uncertainties between the CC/NC ratio and $\sin^2\theta_\odot$
we are unable to estimate the uncertainty on $f_2$ here.
Note, that if the fraction of $\nu_2$ were 100\%,
then $\frac{\text{CC}}{\text{NC}}=\sin^2\theta_\odot$.

\begin{figure}[t]
\centering\leavevmode
\vglue -0.5cm
\includegraphics[width=12.cm]{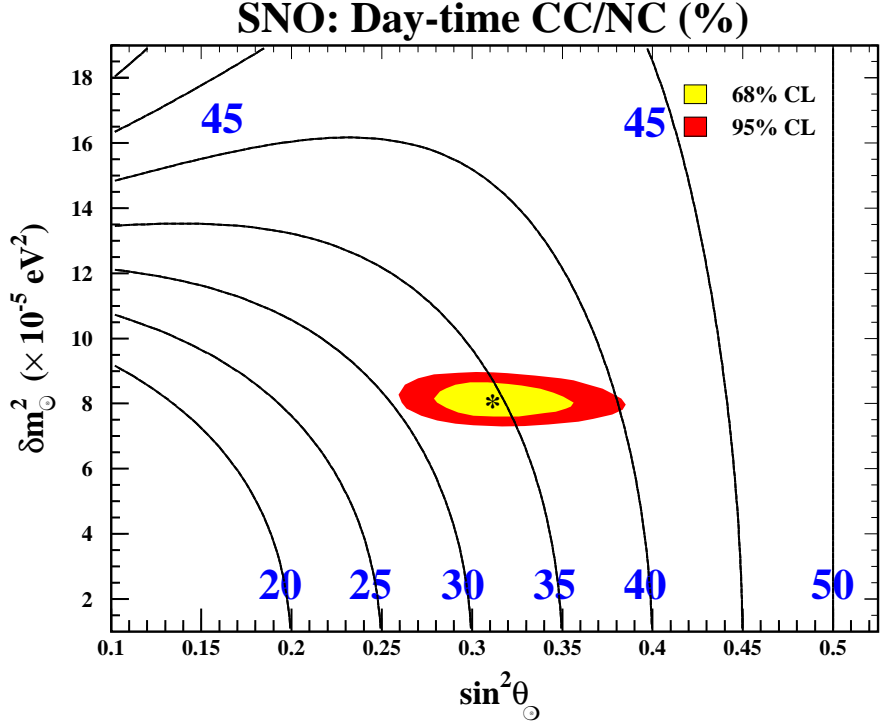}
\vglue -2cm
\caption{SNO's Day-time CC/NC ratio in the $\delta m^2_\odot$ versus
  $\sin^2 \theta_\odot$ plane. At small values of $\delta m^2_\odot$,
  the Day-time CC/NC ratio equals $\sin^2 \theta_\odot$.  The current
  allowed region at 68 and 95\% CL from the combined fit of KamLAND and solar neutrino
  data~\cite{SNO} are also shown by the shaded areas with the
  best fit indicated by the star.  }
\label{cctoncfig}
\end{figure}

Alternatively, we can rewrite Eq.~(\ref{cctonc}) as\footnote{ The
  relationship between day-time $\frac{\text{CC}}{\text{NC}}$ and
  $\theta_\odot$ $\left( =\arcsin \sqrt{ (
      \frac{\text{CC}}{\text{NC}}-f_1)/(1-2f_1)} \right)$ or $\tan^2
  \theta_\odot$ $\left(=( \frac{\text{CC}}{\text{NC}}-f_1)/
    (1-f_1-\frac{\text{CC}}{\text{NC}})\right)$ is not as transparent
  as $\sin^2 \theta_\odot$.}
\begin{eqnarray}
\sin^2 \theta_\odot & = & 
\frac{1}{1-2f_1} \left(\frac{\text{CC}}{\text{NC}} - f_1\right).
\label{sinsq}
\end{eqnarray}
Thus how much CC/NC differs from $\sin^2\theta_\odot$ is
determined by how much $f_2$ differs from 100\%, i.e. the size of
$f_1$.  
In Fig.~\ref{cctoncfig} we have plotted the contours of the
day-time CC/NC ratio in the 
$\sin^2 \theta_\odot$ versus $\delta m^2_\odot$ plane
for the LMA region.  Clearly, at smaller values
of $\delta m^2_\odot$ the day time CC/NC tracks $\sin^2
\theta_\odot$ whereas at larger values an appreciable difference
appears.  This difference is caused by a decrease (increase) in the
fraction that is $\nu_2$ ($\nu_1$) as $\delta m^2_\odot$ gets larger.
Hence if we know the $\nu_1$ or $\nu_2$ fraction we can easily
calculate $\sin^2 \theta_\odot$ from Eq.~(\ref{sinsq}) using a
measured value of the day-time CC/NC ratio.

A similar analysis can also be performed  using the event rate of 
the elastic scattering (ES) at SK and/or 
at SNO. In fact, ES is related to 
the $\nu_1$ and $\nu_2$ fractions, as follows,
\begin{equation}
\frac{\text{ES}}{\text{NC}} = f_1 ( \cos^2\theta_\odot + r \sin^2\theta_\odot)
+f_2 ( \sin^2\theta_\odot + r \cos^2\theta_\odot)
\label{sk_es}
\end{equation}
where $r \equiv \langle \sigma_{\nu_{\mu,\tau} e} 
\rangle /\langle \sigma_{\nu_e e} \rangle \approx$ 0.155
is the ratio of the ES cross sections for 
$\nu_{\mu,\tau}$ and $\nu_e$~\cite{cross_es}, 
averaged over the observed neutrino spectrum. 
Note that we are normalizing  the ES event
rate to that of SNO NC such that Eq. (\ref{sk_es}) is valid 
independent of the SSM prediction of the $^8$B neutrino flux. 

In general, in the presence of neutrino flavor transitions, 
the fraction of $\nu_1$ and $\nu_2$ are not the same for 
ES and CC because the energy dependence of 
the cross sections are different. 
However, in Ref.~\cite{VFL98}, it was suggested that 
if we set analysis threshold energies for SK and SNO
appropriately as
$T_{\text{SNO}} = 0.995 ~T_{\text{SK}} - 1.71 \ (\text{MeV})$, 
where $T_{\text{SNO}}$ and $T_{\text{SK}}$ are 
the kinetic energy threshold of the resulting electron, 
the energy response of these detectors become practically 
identical~\cite{VFL98}. 
Thus, using such a set of thresholds, even if there is a spectral distortion in the recoil electron
energy spectrum, to a good approximation, 
SK/SNO ES and SNO CC are related as follows,
\begin{equation}
\frac{\text{ES}}{\text{NC}} = 
\frac{\text{CC}}{\text{NC}} 
+ r \left(1-\frac{\text{CC}}{\text{NC}}\right),
\label{sk_es2}
\end{equation}
and all the results we obtained for SNO in this paper 
are equally valid for ES at SK and/or at SNO provided the energy thresholds are
set appropriately\footnote{In fact this suggest an alternative to looking for a spectral distortion to test MSW, compare ES to (1-r) CC + rNC for a variety of kinetic energy thresholds.}.

\begin{figure}[t]
\centering\leavevmode
\includegraphics[width=15.cm]{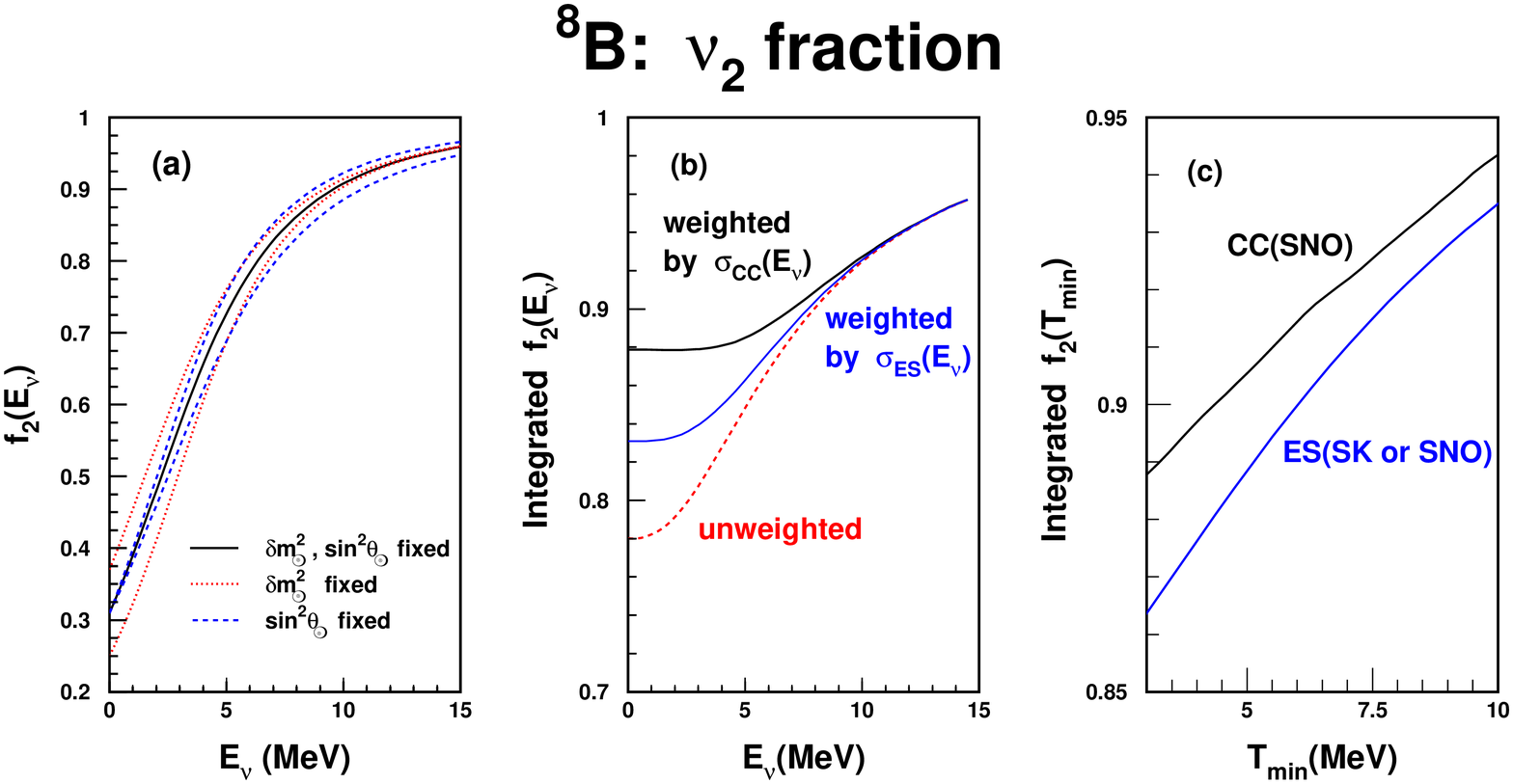}
\vglue -0.8cm
\caption{(a) The fraction of $\nu_2$, $f_2(E_\nu)$,  as a function of
the neutrino energy. The solid (black) curve is obtained using the central
values for $\delta m^2_\odot=8.0 \times 10^{-5}~{\rm eV}^2$ and
$\sin^2 \theta_\odot=0.31$ whereas the blue dashed (red dotted) lines
are the 90\% CL range varying $\delta m^2_\odot$ ($\sin^2
\theta_\odot$) but holding $\sin^2 \theta_\odot$ ($\delta m^2_\odot$)
fixed at the central value, Eq.~(\ref{solar}).
(b) The integrated fraction of $^8$B neutrinos which are $\nu_2$'s above an energy,
$E_\nu$, dashed (red) curve. 
Whereas, the solid black and blue curves are 
weighted by the energy dependence of the charge current (CC) cross
section~\cite{x-sec} and the elastic scattering  (ES) 
cross section~\cite{cross_es}, respectively. 
(c) The integrated fraction of $^8$B neutrinos as a 
function of the threshold kinetic energy of the recoil electrons
for CC (SNO) and ES (SK or SNO) reactions. 
}
\label{f2E}
\end{figure}

In Fig.~\ref{f2E}(a) we show the $\nu_2$ fraction, $f_2(E_\nu)$, versus $E_\nu$. 
The rapid decrease in the $\nu_2$ fraction below $E_\nu \sim 8$ MeV is 
responsible for the expected spectral distortion at energies near threshold in 
both SNO (see Fig. 36 of Ref.~\cite{SNO}) and SK 
(see Fig. 51 of the last Ref. in \cite{SK_solar}).
For a neutrino energy near 10 MeV, the SNO sweet spot, the 90\%
CL variation in $\delta m^2_\odot$ changes $f_2(E_\nu)$ more than
the 90\% CL variation in $\sin^2 \theta_\odot$.  
Whereas in Fig.~\ref{f2E}(b) we give the fraction of $\nu_2$'s 
above a given energy both unweighted and weighted by the energy dependence of 
the CC interaction and ES cross sections.
Note, that above a neutrino
energy of 7.5 MeV there is little difference between the weighted and
unweighted integrated $\nu_2$ fraction.  
Furthermore, in Fig.~\ref{f2E}(c), we show the fraction of $\nu_2$'s 
above a given kinetic energy for the recoil electron 
for both CC (SNO) and ES (SK or SNO) reactions. 
We observe that for the same threshold, $f_2$ for ES is always smaller 
than that for  CC. This is expected since 
unweighted $f_2$ is a increasing function of $E_\nu$ and 
CC cross section increase more rapidly with energy than 
that of ES cross section. 
Hereafter, unless otherwise stated, we focus on the SNO CC reaction,
as the results for  ES reaction are qualitatively similar and the thresholds 
can be adjusted to give identical results for all practical purposes.

\begin{figure}[t]
\centering\leavevmode
\vglue -0.5cm
\includegraphics[width=11.cm]{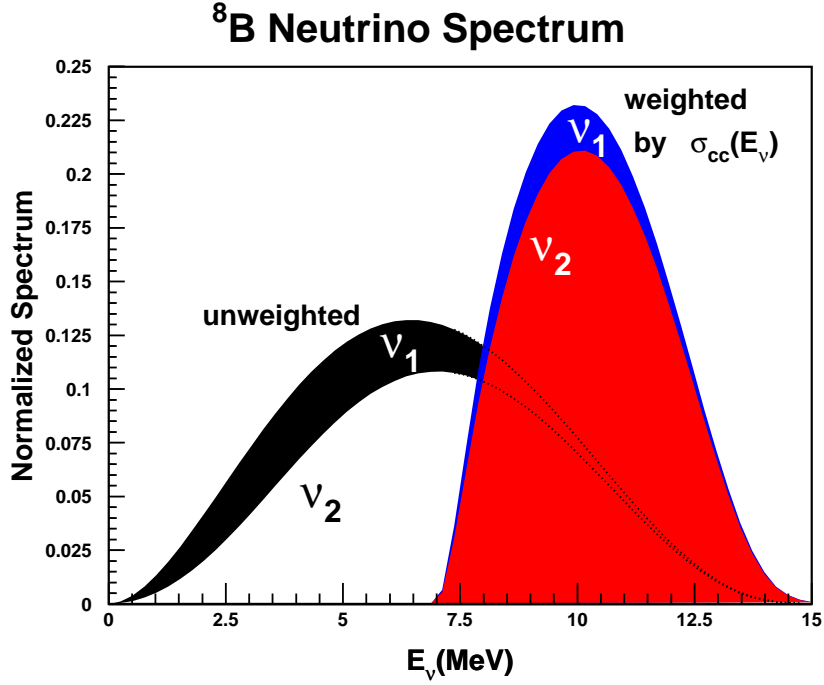}
\vglue -2.0cm
\caption{The normalized $^8$B energy spectrum broken into the $\nu_1$
  and $\nu_2$ components. The left hand curves (black and white) are
  unweighted whereas the right hand curves (blue and red) are weighted
  by the energy dependence of the CC cross
  section~\cite{x-sec} with a threshold of 5.5 MeV for the recoil electron's kinetic energy.
  }
\label{B8_spt}
\end{figure}

In Fig.~\ref{B8_spt} we give
the breakdown into $\nu_1$ and $\nu_2$ for the raw $^8$B spectrum as
well as the spectrum weighted by the energy dependence of the 
CC interaction using a threshold of 5.5 MeV for the kinetic energy of the recoil electrons.  
Here we have used the current best fit values
for $\delta m^2_\odot$ and $\sin^2 \theta_\odot$.

\begin{figure}[htb]
\centering\leavevmode
\includegraphics[width=18.cm]{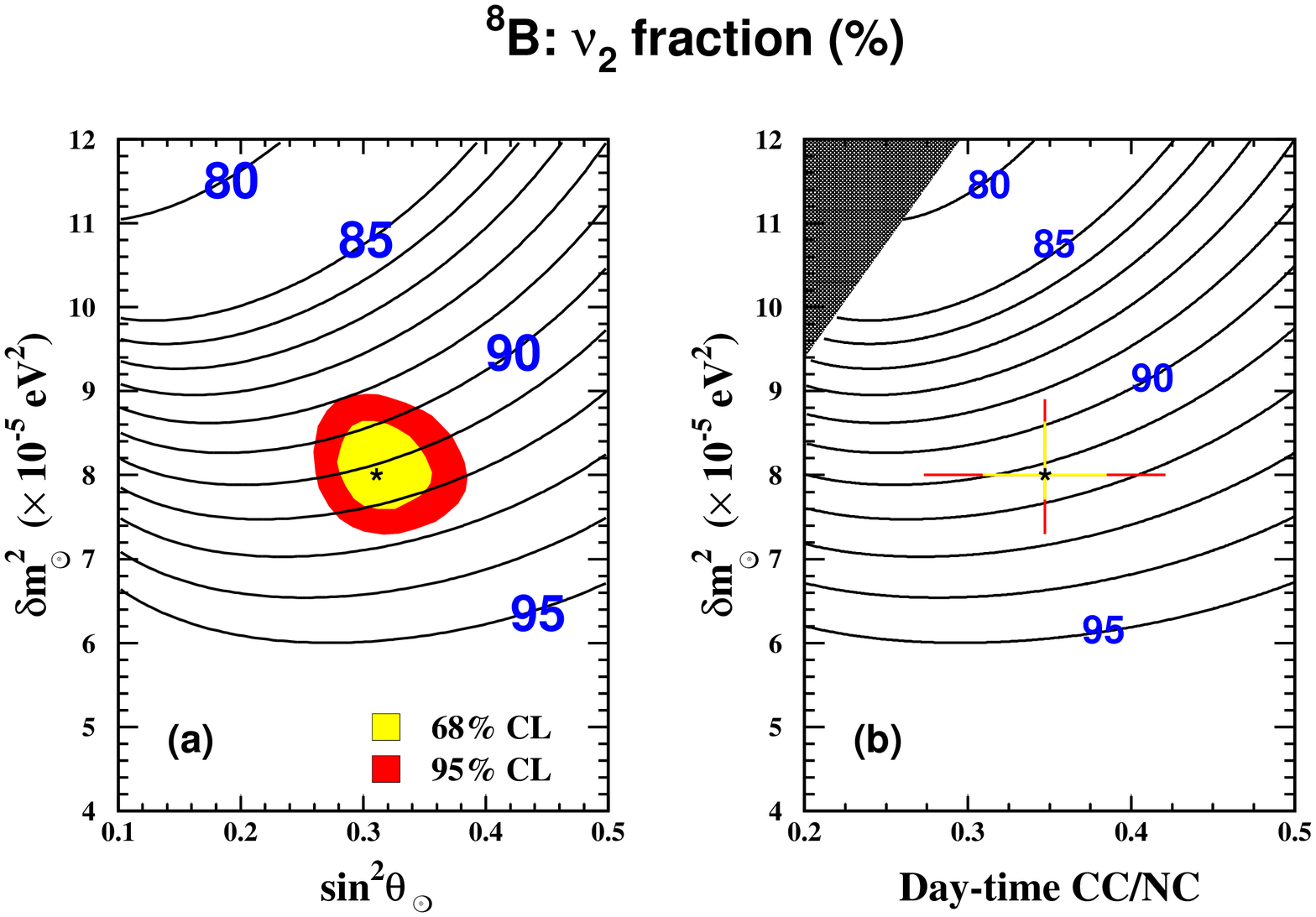}
\vglue -0.5cm
\caption{
(a) The $\nu_2$ fraction (\%) in the $\delta m^2_\odot$ versus $\sin^2
\theta_\odot$ plane. As in Fig.~\ref{cctoncfig}, the current allowed
region is also shown. 
(b) The $\nu_2$ fraction (\%) in the $\delta m^2_\odot$ versus the
Day-time CC/NC ratio of SNO plane. We have excluded a region in
the top left hand corner of this plot which corresponds to
$\sin^2\theta_{\odot}<0.1$. The current allowed range is indicated by
the cross.}
\label{nu2contour}
\end{figure}

How does the fraction of $\nu_2$ vary if we allow $\delta m^2_\odot$
and $\sin^2 \theta_\odot$ to deviate from their best fit values?  In
Fig.~\ref{nu2contour}(a) we show the contours of the fraction of
$\nu_2$ in the $\delta m^2_\odot$ versus $\sin^2 \theta_\odot$ plane
where we have weighted the spectrum by the energy dependence of the
CC interaction cross-section, and we have used a threshold on the kinetic energy
of the recoil electrons of 5.5 MeV.  This energy dependence mimics the energy
dependence of the SNO detector.  Because of the strong correlation
between $\sin^2 \theta_\odot$ and the day-time CC/NC ratio we also
give the contours of the fraction of $\nu_2$ in the $\delta m^2_\odot$
versus day-time CC/NC plane in Fig.~\ref{nu2contour}(b).  
Thus the $^8$B energy weighted average fraction of $\nu_2$'s observed 
by SNO is 
\begin{equation}
f_2 = 91\pm2\% \quad \text{at the 95\% CL.}
\end{equation}
This is the two neutrino answer to the question posed in the title of this paper.
We note, however, that as we showed in Fig.~\ref{f2E}(c) the value of $f_2$ is a function 
of the threshold energy and also depends on the experiment. We estimate that
for SK with the current 4.5 MeV threshold for the kinetic energy of the recoil electrons,
that
\begin{equation}
f_2 = 88 \pm 2\% \quad \text{at the 95\% CL.}
\end{equation}
The uncertainty is dominated by the uncertainty in $\delta
m^2_\odot/A$.   However, the uncertainty on $\delta m^2_\odot$ 
is approximately 5\% from the KamLAND data whereas the uncertainty on
the matter potential, $A$,  in the region of $^8$B production of the Standard Solar Model  is
1-2\%, see \cite{BPB2001}.  Hence, the uncertainty on $\delta m^2_\odot$ dominates.

For the current allowed values for $\delta m^2_\odot$ and $\sin^2 \theta_\odot$,
the ratio
\begin{eqnarray}
 \frac{\delta m^2_{\odot} \sin 2 \theta_{\odot} }
{A(^8\text{B})-\delta m^2_{\odot} \cos 2 \theta_{\odot}}
&  \approx & \frac{3}{4},  
\label{8Bapprx1}
\end{eqnarray}
where $A(^8\text{B})$ is obtained using a typical number density of
electrons at $^8$B neutrino production ($Y_e \rho \approx 90 {~\rm
  g.cm^{-3}}$) and the typical energy of the observed $^8$B neutrinos
($\approx$ 10 MeV).
 
For the best fit central values of $\delta m^2_\odot$
and $\sin^2 \theta_\odot$, given by Eq.(\ref{solar}), let us define an 
effective matter potential for the $^8$B  neutrinos, $A^{^8\text{B}}_{eff}$,
such that the left hand side of Eq.(\ref{sinsqthN}) equals
our best fit value for the fraction that is $\nu_2$.     
Thus,
\begin{eqnarray}
A^{^8\text{B}}_{eff} & \equiv &
\delta m_\odot^2 \sin 2\theta_\odot 
\left[ 
\cot 2\theta_\odot + \frac{2f_2-1}{2\sqrt{f_2(1-f_2)}} \right] 
\label{Aeff} \\
 &=& 1.36 \times 10^{-4} ~{\rm eV}^2, 
  \nonumber
\end{eqnarray}
for $f_2=0.910$.
This $A^{^8\text{B}}_{eff}$ corresponds to a $Y_e\rho E_\nu = 0.892 ~\text{kg cm}^{-3} ~\text{MeV}$, 
the effective mixing angle, $\theta_\odot^N\vert_{eff}= 73^\circ$ and the effective
$\delta m^2_N\vert_{eff}=13.6 \times 10^{-5}$ eV$^2$.

We can then use this $A^{^8\text{B}}_{eff}$ to perform a
Taylor series expansion about the best fit point as follows
\begin{eqnarray}
f_2  =  \langle \sin^2 \theta^N_\odot  \rangle_{^8\text{B}} \approx 
 \frac{9}{10} + \frac{24}{125} \xi 
 + {\cal O}(\xi^2) \quad  
&  {\rm with}   & \xi  \equiv   \frac{3}{4} -\frac{\delta m^2_{\odot} 
\sin 2 \theta_{\odot} }{(A^{^8\text{B}}_{eff} -\delta m^2_{\odot} \cos 2 \theta_{\odot})}.
\label{8Bapprx}
\end{eqnarray}
This simple expression reproduces the values of $f_2$ to high
precision throughout the 95\% allowed region of the KamLAND and 
the solar neutrino experiments given in Fig.~\ref{nu2contour}(a).
In this sense our $A^{^8\text{B}}_{eff}$ is the effective matter potential
for the $^8$B neutrinos.
An expansion in $\delta m^2_{\odot}/A$ around its typical value of $0.6$ could also be used
but the coefficients are ever more complex trigonometric functions of  $\theta_\odot$, whereas
with our $\xi$ expansion the coefficients are small rational numbers.

\subsection{$^7$Be and pp neutrinos}

For $^7$Be and pp neutrinos the fractions of $\nu_1$ and $\nu_2$ are
much closer to the vacuum values of $\cos^2 \theta_\odot$ and $\sin^2
\theta_\odot$ respectively, as they are produced well below (more than a factor of two) the 
MS-resonance in the Sun, 
and an expansion in $A/\delta m^2_{\odot}$
is the natural one.  
In the third Ref. in ~\cite{pxzero}, the electron neutrino survival probability was obtained by a similar expansion around the average of the matter potential. 
Using this expansion, we find that
\begin{eqnarray}
f_2 =  1 -f_1=\sin^2 \theta^N_\odot   
= \sin^2 \theta_\odot  & + & 
\frac{1}{2} \sin^2 2\theta_\odot \left(\frac{A}{\delta m^2_{\odot} } 
\right) +
{\cal O}\left(\frac{A}{\delta m^2_{\odot} }\right)^2  \label{otherapprx}\\[0.3cm]
{\rm with} \quad A^{^7\text{Be}}_{eff}= 1.1\times 10^{-5} {\rm eV}^2  & &
{\rm and } \quad A^{\text{pp}}_{eff} = 0.31 \times 10^{-5} {\rm eV}^2, 
\end{eqnarray}
where the averaged value of the energy (weighted by the cross section)
as well as the electron densities used are, respectively, 
$\langle E_\nu \rangle_{\text{pp}}$ = 0.33 MeV and $\langle Y_e \rho
\rangle_{\text{pp}}$ = 62 g/cm$^3$ for pp, and 
$\langle E_\nu \rangle_{^7\text{Be}}$ = 0.86 MeV 
and $\langle Y_e \rho \rangle_{^7\text{Be}}$ 
= 81 g/cm$^3$ for $^7$Be.
Thus $f_2(^7\text{Be}) = 37 \pm 4 (7) \%$ and $f_2(\text{pp}) = 33 \pm
4 (7)\%$ at 68 (95) \% CL where the uncertainty 
here is dominated by our knowledge of $\sin^2\theta_\odot$.  

\begin{figure}[htb]
\centering\leavevmode
\vglue -1.0cm
\hglue 1.0cm
\includegraphics[width=15.cm]{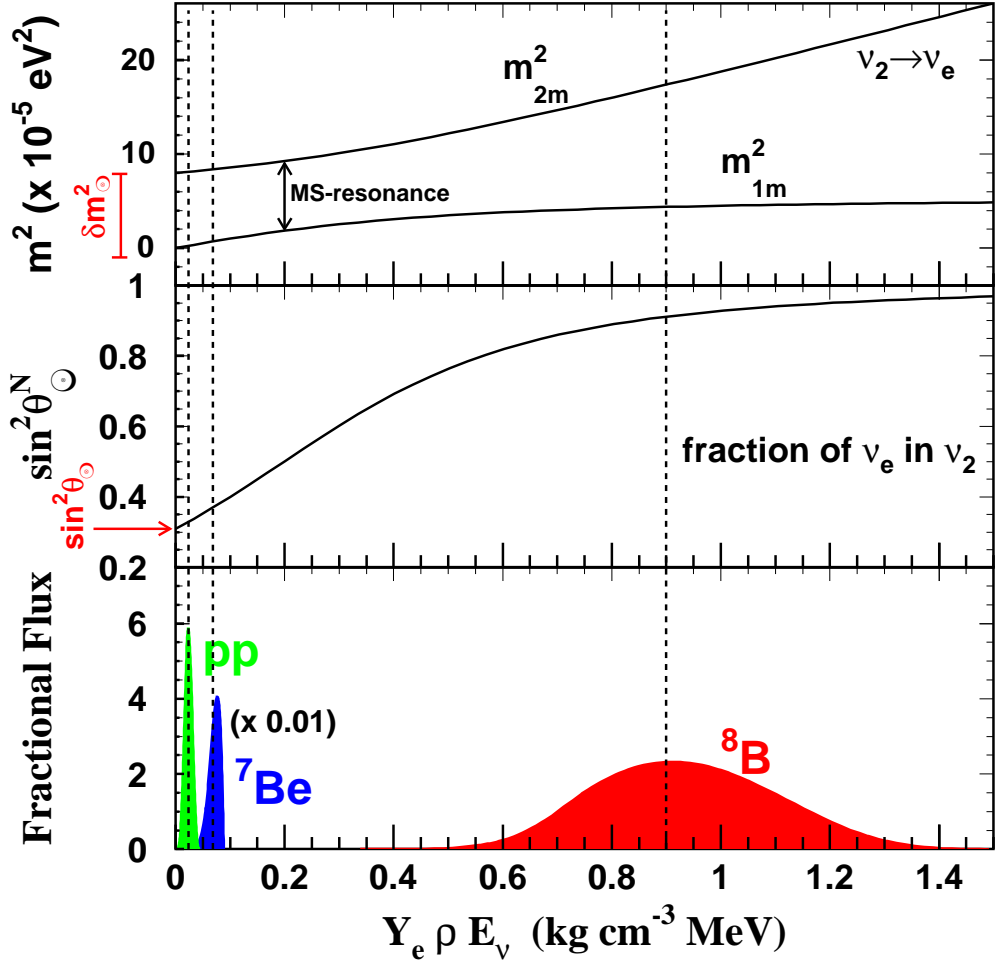}
\vglue -0.8cm
\caption{The Mass spectrum (top panel), the fraction of
  $\nu_2$'s produced, $\sin^2\theta_{\odot}^{N}$, (middle panel) and the
  fractional flux (bottom panel) versus the product of the electron
  fraction, $Y_e$, the matter density, $\rho$, and the neutrino energy,
  $E_\nu$, for the best fit values $\delta m^2_{\odot}=8.0\times 10^{-5} \text{eV}^2$ and $\sin^2\theta_{\odot}=0.310$.
  The vertical dashed lines give the value of
  $Y_e \rho E_\nu$ which reproduces the average $\nu_2$ fractions, 91, 37 and 33\%
  for $^8$B, $^7$Be and pp respectively. 
  This value of $Y_e \rho E_\nu=0.89$ kg cm$^{-3}$ MeV,  for the $^8$B neutrinos, 
  gives a production mixing angle equal to 73$^\circ$ and a production
  $\delta m^2_N =14\times 10^{-5}$ eV$^2$.
  $Y_e \rho E_\nu = 1$ kg cm$^{-3}$ MeV corresponds, in terms of the matter potential,  
  to $15.3 \times 10^{-5}$ eV$^2$, see Eq. (\ref{Adefn}).  }
\label{mass-flux}
\end{figure}

\subsection{Two Neutrino Summary}

In Fig.~\ref{mass-flux} we give the neutrino mass spectrum, the value
of fraction of $\nu_2$'s ($\sin ^2 \theta_\odot^N $) and the
fractional flux as function of
the electron number density times neutrino energy,
$Y_e \rho E_\nu $, which 
is proportional to the matter potential, 
for the $^8$B, $^7$Be and pp neutrinos using the best fit
values of $\delta m^2_\odot$ and $\sin^2 \theta_\odot$ in Eq.~(\ref{solar}).  
The $^8$B energy spectrum has been weighted by the
energy dependence of the CC interaction of SNO with a 5.5
MeV threshold on the kinetic energy of the recoil electrons
whereas the pp energy spectrum has been weighted by the
energy dependence of the charged current interaction on Gallium with a
0.24 MeV threshold.  The vertical dashed lines gives the value of
$Y_e \rho E_\nu$ which reproduces the average $\nu_2$ fraction
using the
simple expression in Eq.~(\ref{sinsqthN}) and are useful for the
approximations given in Eqs.~(\ref{8Bapprx}) and (\ref{otherapprx}).

The energy weighted $\nu_2$ fractions for $^8$B, $^7$Be and pp
neutrinos using a two neutrino analysis, at the 95\% CL, are
\begin{eqnarray}
f_2(^8\text{B}) & = & 91 \pm 2\%, \\
f_2(^7\text{Be}) &= & 37 \pm 4\%, \\
f_2(\text{pp}) & = & 33 \pm 4\%,
\end{eqnarray}
where the uncertainties for $^7$Be and pp are dominated by the
uncertainty on $\sin^2\theta_\odot$ whereas for $^8$B the uncertainty
is dominated by the uncertainty on $\delta m^2_{\odot}$. The $\nu_1$
fractions, $f_1$, are simply $1-f_2$.

\section{Three Neutrino Analysis}

For the three neutrino analysis we first must discuss the size of the
component of $\nu_3$ which is $\nu_e$, i.e. the size of $\sin^2
\theta_{13}$.  This mixing angle determines the size of the effects on $\nu_e$ 
associated with the atmospheric mass squared difference.  The best constraint on 
$\theta_{13}$
comes from
the CHOOZ reactor experiment ~\cite{Chooz} which gives a limit on
$\sin^2\theta_{13}$, as
\begin{eqnarray}
0 \leq & \sin^2 \theta_{13} & < 0.04, 
\end{eqnarray}
at the 90 \% CL for
$\delta m^2_{31} = 2.5 \times 10^{-3} {\rm eV^2}$. 
This constraint depends on the precise value of
$\delta m^2_{31}$ with a stronger (weaker) constraint at higher
(lower) allowed values of $\delta m^2_{31}$.

So far the inclusion of genuine three flavor effects has not been
important because these effects are controlled by the two small
parameters
\begin{eqnarray}
{\displaystyle\delta m^2_{21} \over \displaystyle \delta m^2_{32}}\approx 0.03
& \quad {\rm and/or} \quad  \sin^2 \theta_{13} \leq 0.04.
\end{eqnarray}
However as the accuracy of the neutrino data improves it will become
inevitable to take into account genuine three flavor effects. 
See ~\cite{Fogli05,smir+gos}, for recent studies 
on the impact of $\theta_{13}$ on solar neutrinos.
 
Suppose that Double CHOOZ~\cite{DoubleC}, T2K~\cite{T2K} or
NO$\nu$A~\cite{NOvA} or some other experiment measures a non-zero
value for $\sin^2 \theta_{13}$.  What effect does this have on the
previous analysis?  How does this change our knowledge of the solar
parameters and the relationship between solar mixing angle and the
fraction of $\nu_2$?

Our knowledge of the solar $\delta m^2$ comes primarily from the
KamLAND experiment where the effects of the atmospheric $\delta m^2$
are averaged over many oscillations, thus to high accuracy
\begin{eqnarray}
\delta m^2_{21} & = & \delta m^2_\odot,
\end{eqnarray}
i.e. the solar $\delta m^2$ remains unaffected.  Remember, we are using the
notation $\delta m^2_{21}$ and $\sin^2\theta_{12}$ for the three
neutrino analysis to distinguish it from $\delta m^2_\odot$ and
$\sin^2\theta_\odot$ used in the two neutrino analysis.

\subsection{$^8$B 3 Neutrino Analysis}

For the mixing angle $\sin^2 \theta_{12}$ the situations is more
complicated in the three neutrino analysis.  The $^8$B electron
neutrino survival probability measured by SNO's day-time CC/NC ratio
can be written as
\begin{eqnarray}
\frac{\text{CC}}{\text{NC}}  & = & \mathcal{F}_1 \cos^2 \theta_{13} \cos^2 \theta_{12} 
                             +\mathcal{F}_2 \cos^2 \theta_{13} \sin^2 \theta_{12}
			  +\mathcal{F}_3 \sin^2\theta_{13},
			  \label{ccovernc3}
\end{eqnarray}
\newcommand{\f}{\mathcal{F}} where $\f_1$, $\f_2$ and $\f_3$ are the
fraction of $\nu_1$, $\nu_2$ and $\nu_3$ respectively, satisfying
$\f_1+\f_2+\f_3=1$.  The $\nu_3$ fraction is given by
\begin{equation}
\f_3 = \left(1\pm \frac{2A}{|\delta m^2_{31}|}\right)\sin^2\theta_{13}\approx \sin^2\theta_{13},
\end{equation}
where +(-) sign refers to the normal, $\delta m^2_{31} >0$ (inverted,
$\delta m^2_{31} <0$) mass hierarchy.  The small correction factor
$\frac{2A}{|\delta m^2_{31}|} \sim 10\%$ comes from matter effects
associated with atmospheric $\delta m^2$ in the center of the Sun. We
will ignore this correction since it is small and currently the sign
is unknown.
Hence, $\f_1+\f_2 =1-\f_3= \cos^2\theta_{13}$.

With this approximation the $\nu_1$ and $\nu_2$ fractions can be
written as
\begin{eqnarray}
\f_1=\cos^2\theta_{13}~\langle  \cos^2\theta^N_{12}  \rangle_{^8\text{B}} \quad & {\rm and}  & \quad
  \f_2=\cos^2\theta_{13}~\langle  \sin^2\theta^N_{12}  \rangle_{^8\text{B}}.
\label{ff1}
\end{eqnarray}
where the average $\langle \cdots \rangle_{^8\text{B}}$ is over the
solar production region and the energy of the observed neutrinos.
$\sin^2\theta^N_{12} $ is given by Eq.~(\ref{sinsqthN}) with the
replacements $\sin^2 \theta_\odot \rightarrow \sin^2 \theta_{12}$ and
$A \rightarrow A\cos^2\theta_{13}$~\cite{Lim_3g}.

\begin{figure}[htb]
\centering\leavevmode
\vglue -.5cm
\includegraphics[width=15.cm]{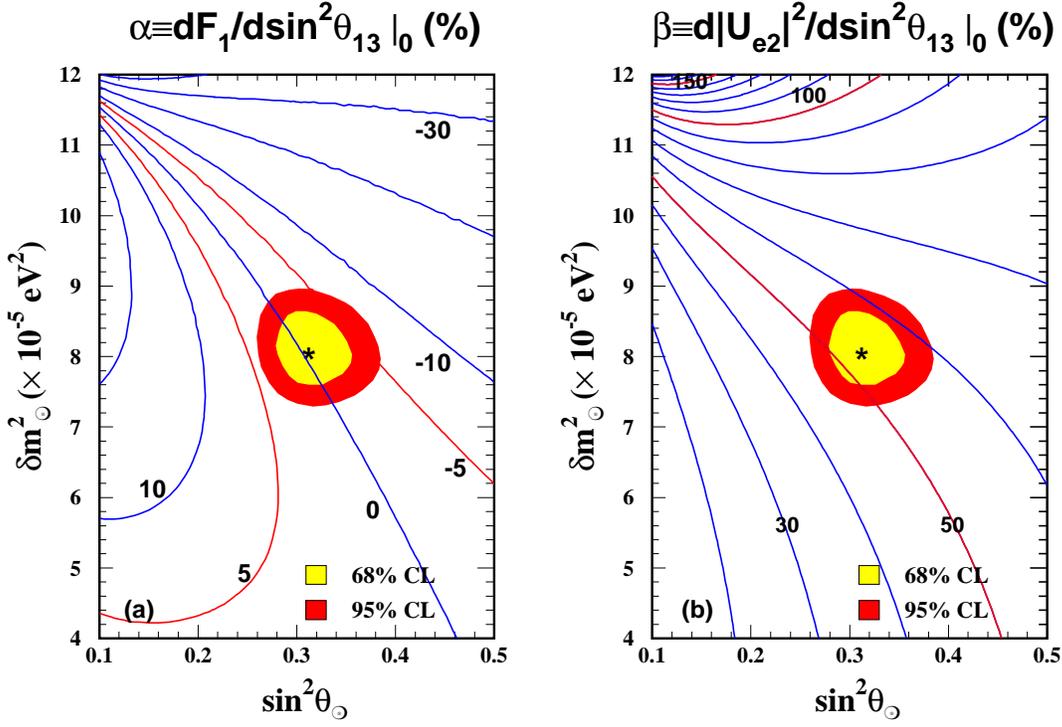}
\vglue -0.5cm
\caption{Iso-contours of the derivatives of $\f_1$ 
(a) and $\vert U_{e2}\vert^2$ 
(b) with respect $\sin^2 \theta_{13}$ evaluated at $\sin^2 \theta_{13}=0$
in the $\delta m^2_{\odot}$ versus $\sin^2\theta_{\odot}$ plane.  The contours are labeled as
in per cent.
The 68 and 95 \% CL allowed regions are also indicated.
}
\label{alpha}
\end{figure}

In going from the two neutrino analysis to the three neutrino analysis
the quantity that must remain unchanged is the value of the electron
neutrino survival probability, i.e. the CC/NC ratio.  This implies
that we must adjust the value of $\sin^2 \theta_{12}$ and hence the
fractions of $\nu_1$ and $\nu_2$ so that the CC/NC ratio remains
constant.  We have performed this procedure numerically and report the
result as a Taylor series expansion in the fraction of $\nu_1$'s about
$\sin^2 \theta_{13}=0$. 
If we write
\begin{eqnarray}
  \f_1(\sin^2\theta_{13}) & = & 
\f_1(0) + \alpha \sin^2\theta_{13}  + {\cal O}( \sin^4\theta_{13} ), \\
  {\rm then} \quad \f_1(0)\equiv f_1, \quad & {\rm and}&  \quad \alpha  \equiv  \left. \frac{d\f_1}{d\sin^2 \theta_{13}}\right|_{\sin^2 \theta_{13} = 0}. 
\end{eqnarray}
In Fig.\ref{alpha}(a) we have plotted the contours of $ \alpha \equiv \left.
  \frac{d\f_1}{d\sin^2 \theta_{13}}\right|_{ 0} $ in the $\delta
m^2_\odot$ versus $\sin^2\theta_\odot$ plane.  Near the best values this
total derivative is close to zero, i.e.
\begin{equation}
\left. \frac{d\f_1}{d\sin^2 \theta_{13}}\right|_{\sin^2 \theta_{13} = 0} 
= 0.00^{+0.02}_{-0.04}  
\end{equation}
at the 68\% CL.
As $\sin^2 \theta_{13}$ grows above zero, the size of $\f_1$ is
influenced by a number of effects; the first is the factor of $\cos^2
\theta_{13}$ in Eq.~(\ref{ff1}) which reduces $\f_1$, the second is
the matter potential A which is reduced to $A\cos^2 \theta_{13}$  raising 
the fraction $\f_1$ and third is the value of $\sin^2 \theta_{12}$ which 
changes to hold the CC/NC ratio fixed.  By coincidence the sum of
these effects approximately cancel at the current best fit values and
the fraction of $\nu_1$ remains approximately {\it unchanged} as
$\sin^2 \theta_{13}$ gets larger. This implies that the fraction of
$\nu_2$ is reduced by $\sim \sin^2 \theta_{13}$ since the sum of $\f_1+\f_2$ is
simply $\cos^2 \theta_{13}$, thus
\begin{eqnarray}
\f_1 & \approx&  f_1   = 0.09 \mp 0.02,  \\
\f_2  & = & f_2 - \sin^2 \theta_{13} 
\approx  0.91\pm 0.02 -\sin^2\theta_{13}, \\
\f_3 & = & \sin^2 \theta_{13}.
\end{eqnarray}
Remember $f_i$ and $\f_i$ are the fractions of the $i$-th mass eigenstate in the two and three neutrino analysis, respectively.
The uncertainty comes primarily from the uncertainty in $\delta m^2_\odot$ measured by KamLAND.

As a use of these fractions one can for example evaluate the MNS
matrix element, $\vert U_{e2}\vert^2 =
\cos^2\theta_{13}\sin^2\theta_{12}$, by rewriting
Eq.~(\ref{ccovernc3}) as
\begin{eqnarray}
\vert U_{e2}\vert^2 = \cos^2\theta_{13}\sin^2\theta_{12} & = &
\frac{(\frac{\text{CC}}{\text{NC}} - \cos^2\theta_{13} \f_1)} 
{(\cos^2 \theta_{13} - 2\f_1)} , 
\end{eqnarray}
where terms of ${\cal O}(\sin^4 \theta_{13})$ have been dropped.
Performing a Taylor series expansion about $\sin^2 \theta_{13}=0$, we
find
\begin{eqnarray}
\vert U_{e2}\vert^2 &=& \sin^2\theta_\odot ^{^8\text{B}}  + \beta \sin^2 \theta_{13}
+{\cal O}(\sin^4\theta_{13}), \\[0.25cm]
{\rm with}\quad 
\beta & \equiv &  \left. \frac{d \vert U_{e2}\vert^2}{d \sin^2\theta_{13}} \right|_0   = 
\frac{(f_1-\alpha)+(1+2\alpha) \sin^2\theta_\odot }
{(1-2f_1)}.
\end{eqnarray}		  
For the current allowed region of the solar parameters, this implies
that
\begin{eqnarray}
\vert U_{e2} \vert^2 
& \approx & \sin^2\theta_\odot ^{^8\text{B}}   + (0.53 ^{+0.06}_{-0.04})\sin^2 \theta_{13},
\label{U8B}
\end{eqnarray}
at the 68\% CL,
i.e. the three neutrino $\vert U_{e2}\vert^2$ is approximately equal
to the $\sin^2\theta_\odot^{^8\text{B}}  $ 
using a two neutrino analysis of only the $^8$B
electron neutrino survival probability using the KamLAND's $\delta m^2_\odot$ constraint 
plus 53\% of $\vert
U_{e3}\vert^2$ determined, say, by a CHOOZ-like reactor experiment, see 
Fig.~\ref{alpha}(b).

If a similar analysis is performed for the three neutrino sine squared
solar mixing angle $\sin^2 \theta_{12}$, the total derivative with
respect to $\sin^2 \theta_{13}$ is simply
$(\beta+\sin^2\theta_\odot)$.  For $\tan^2 \theta_{12}$ the total
derivative is $(\beta+\sin^2\theta_\odot)/\cos^4\theta_\odot$.
Alternatively we can turn this discussion inside out and write the
$^8$B effective two component $\sin^2 \theta_\odot$ in terms of three
component quantities as
\begin{eqnarray}
\sin^2 \theta_\odot^{^8\text{B}}  & = & \sin^2 \theta_{12}-(\beta+\sin^2 \theta_{12}) \sin^2 \theta_{13}.
\end{eqnarray}
For KamLAND, the equivalent relationship is 
\begin{eqnarray}
\sin^2 \theta_\odot^{^8\text{Kam}}  & = & \sin^2 \theta_{12}
-\left( \frac{\sin^22\theta_{12}}{2\cos 2 \theta_{12}} \right) \sin^2 \theta_{13}.
\end{eqnarray}
For the current best fit values $(\beta+\sin^2 \theta_{12}) \approx 0.90$ is close to
$\sin^22\theta_{12}/2\cos 2 \theta_{12} \approx 1.1$, i.e. in a two component
analysis the difference between the solar $^8$B and KamLAND $\sin^2 \theta_\odot$'s
is approximately $0.2 \sin^2 \theta_{13}$.

\subsection{$^7$Be and pp 3 Neutrino Analysis}

Performing a similar 3 neutrino analysis for the  pp (or $^7$Be)
neutrinos we find that the fraction of neutrino mass eigenstates is
\begin{eqnarray}
\f_1   & \approx & \cos^2 \theta_\odot  
-  \frac{1}{2} \sin^2 2\theta_\odot \left(\frac{A}{\delta m^2_{\odot} } 
\right)  + \frac{\sin^2 \theta_\odot}{\cos 2 \theta_\odot} \sin^2 \theta_{13}  
= f_1 + 0.82 \sin^2 \theta_{13},\\
\f_2 &  \approx & \sin^2 \theta_\odot  
+  \frac{1}{2} \sin^2 2\theta_\odot \left(\frac{A}{\delta m^2_{\odot} } 
\right)  - \frac{\cos^2 \theta_\odot}{\cos 2 \theta_\odot} \sin^2 \theta_{13}
= f_2 - 1.8 \sin^2 \theta_{13}, \\
\f_3  & \approx & \sin^2 \theta_{13},
\label{7be+pp3}
\end{eqnarray}
where the $\sin^2 \theta_\odot$ here is determined from the pp (or $^7$Be) neutrinos.
Terms of order ${\cal O}\left(A/\delta m^2_{\odot} \right)^2$, ${\cal
  O}(\sin^4\theta_{13})$ and ${\cal O}\left(\sin^2\theta_{13} A/\delta
  m^2_{\odot} \right) $ have been dropped here.  The two neutrino
fractions $f_1$ and $f_2$ are given in Eq.~(\ref{otherapprx}).

Again we can use these fractions to determine the $\vert U_{e2}
\vert^2$ element of the MNS matrix
\begin{eqnarray}
\vert U_{e2} \vert^2 & = & \sin^2\theta_\odot - \left( \frac{\cos^2 \theta_\odot}{\cos 2\theta_\odot} \right) \sin^2\theta_{13} 
\approx  \sin^2\theta_\odot - 1.8 ~\sin^2\theta_{13}.
\label{Upp}
\end{eqnarray}
Comparing this equation with Eq.~(\ref{U8B}) appears to be in
contradiction but this is not so since if $\sin^2 \theta_{13} \neq 0$
then the two component analysis of the $^8$B and pp (or $^7$Be)
neutrinos will lead to different values of $\sin^2 \theta_\odot$, in
fact
\begin{eqnarray}
\sin^2\theta_\odot^{\text{pp}} -\sin^2 \theta_\odot^{^8\text{B}} \approx  2.3 \sin^2 \theta_{13}.
\end{eqnarray}
This difference has been extensively exploited in Ref.~\cite{smir+gos}
to determine $\sin^2\theta_{13}$ using only solar neutrino
experiments.  Their $\sin^2 \theta_{12}$ versus $\sin^2 \theta_{13}$
figures, e.g. Fig.~6, demonstrates this point in a clear and useful
fashion.  Also, the numerical values of our derivatives of $\vert
U_{e2} \vert^2$ are consistent with the inverse of the slopes of their
Fig.~6.

Eqs.~(\ref{U8B}) and (\ref{Upp}) also imply that the uncertainty in
the determination of $\vert U_{e2} \vert^2$ from the current unknown
value of $\sin^2 \theta_{13}$ is smaller for the analysis of $^8$B
neutrinos than pp or $^7$Be neutrinos.  Of course the current
uncertainty on the two neutrino $\sin^2 \theta_\odot$ dominates.

\section{Probing the solar interior by $^8$B neutrinos}

In this section, as an application of our analysis, we will 
invert the discussions found in Ref.~\cite{MSW-test}
where the validity of the MSW physics has been tested assuming
the standard solar model (SSM) prediction of the electron number density 
as well as $^8$B neutrino production region.
Here, we will discuss what can be
said about these quantities, assuming the validity of the MSW effect in
the LMA region.  While there is no strong reason to
doubt the correctness of the SSM, 
which is in good agreement also with the helioseismological data~\cite{helio}, 
it is nevertheless interesting if we can test it independently.

Since the propagation of $^8$B neutrinos, in the Sun,
is highly adiabatic in the LMA region, the fraction of $\nu_2$, and
consequently, the SNO CC/NC ratio is determined only by the
effective value of the matter potential, $A^{^8\text{B}}_{eff}$, 
defined in Section II(A).
This implies that if we can measure $\sin^2\theta_\odot$  using an experiment independent
of the $^8$B solar neutrinos, 
then from the measured value of SNO's CC/NC ratio
we can determine the value of $A^{^8\text{B}}_{eff}$.
Note, that we can not extract information
on the electron number density distribution or the $^8$B neutrino production
distribution, separately, but only on $A^{^8\text{B}}_{eff}$ which is a single characteristic
of the convolution of these two distributions.

\begin{figure}[t]
\centering\leavevmode
\vglue -0.5cm
\includegraphics[width=9.6cm]{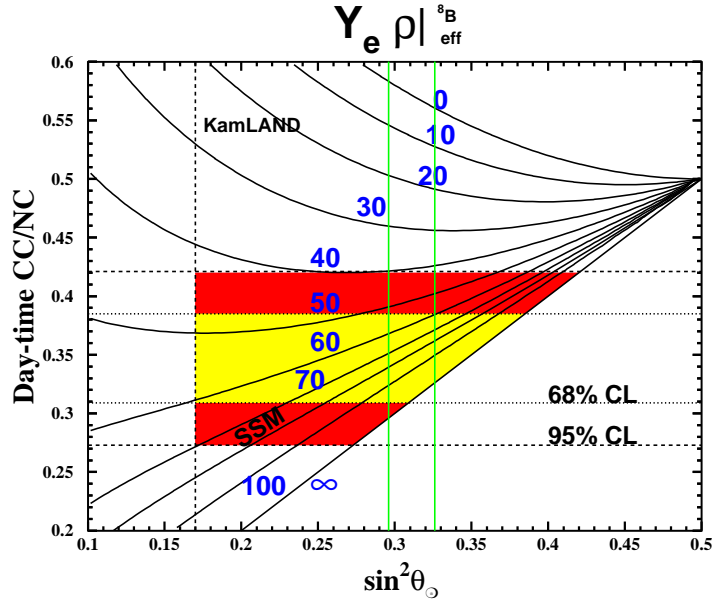}
\vglue -1.6cm
\caption{The iso-contours of $Y_e\rho~\vert^{^8\text{B}}_{eff}$
in the $\sin^2\theta_{\odot}-\text{CC/NC}|_{\text{day}}$ plane.
The line labeled SSM is the Standard Solar Model prediction for
$Y_e\rho~\vert^{^8\text{B}}_{eff}$ ($\approx$ 85 g/cm$^3$).
The range of observed values of CC/NC are indicated by 
the shaded horizontal bands. 
The KamLAND experiment places a lower bound on $\sin^2 \theta_\odot$
independent of solar neutrinos
at 0.17, see \cite{KamLAND}.
The vertical band indicate the uncertainty which 
could be expected by future reactor experiments~\cite{SADO}.  
}
\label{yerho_eff}
\end{figure}

For the two flavor neutrino analysis, if we rewrite the definition of
the effective matter potential $A^{^8\text{B}}_{eff}$ given by Eq.(\ref{Aeff}) using the
relationship between $f_2$ and SNO's CC/NC ratio, Eq.(\ref{f1eqn}), 
we obtain
\begin{eqnarray}
A^{^8\text{B}}_{eff}
&= &
\delta m_\odot^2 \sin 2\theta_\odot 
\left[ 
\cot 2\theta_\odot + \frac{1-2\frac{\text{CC}}{\text{NC}}}
{2\sqrt{(\cos^2\theta_\odot-\frac{\text{CC}}{\text{NC}})
(\frac{\text{CC}}{\text{NC}}-\sin^2\theta_\odot)}} \right].
\label{Aeff-SNO}
\end{eqnarray}
This expression allows us to obtain a value of $A^{^8\text{B}}_{eff}$ from
($\sin^2\theta_\odot$, $\delta m^2_\odot$) measured independent of $^8$B neutrinos
and SNO's $^8$B neutrino CC/NC ratio.
We can convert this into an effective value of the electron  
number density, $Y_e\rho~\vert^{^8\text{B}}_{eff}$,  in the solar $^8$B production region, 
as follows
\begin{equation}
Y_e\rho~\vert^{^8\text{B}}_{eff}
\equiv \displaystyle \frac{M_n}{2\sqrt{2} G_F}
\frac{A^{^8\text{B}}_{eff}}{\langle E_\nu \rangle_{^8\text{B}}}, 
\end{equation}
where $\langle E_\nu\rangle_{^8\text{B}}=10.5$ MeV is the CC
cross section weighted average energy of neutrinos 
observed by SNO.
For a given solar model, the value of $Y_e\rho~\vert^{^8\text{B}}_{eff}$ can be calculated
for any value of $\sin^2\theta_\odot$ and $\delta m^2_\odot$.
The SSM prediction is that
$Y_e\rho~\vert^{^8\text{B}}_{eff} =85 ~\text{g~cm}^{-3}$ at the current best fit point\footnote{Because of the way we have defined $A^{^8\text{B}}_{eff}$, our $Y_e\rho~\vert^{^8\text{B}}_{eff}$ has a weak
dependence on $\sin^2\theta_\odot$ and $\delta m^2_\odot$ but this variation is less than 2\% over the
95\% CL allowed region.}. As a comparison the mean value of $Y_e\rho$ over the $^8$B production
region is 90  g~cm$^{-3}$. The reason that $Y_e\rho~\vert^{^8\text{B}}_{eff} $ is below the mean value 
is because values of $Y_e\rho$  below the mean pull down the $\nu_2$ fraction more than values above the mean raise the $\nu_2$ fraction.

We show in Fig.~\ref{yerho_eff}, the iso-contours of $Y_e\rho~\vert^{^8\text{B}}_{eff}$
 in the
$\sin^2\theta_\odot-\text{CC}/\text{NC}|_{\text{day}}$ plane, for the
current best fitted value of $\delta m_\odot^2$.  The observed range of SNO's
CC/NC are shown by the horizontal lines\footnote{Another horizontal band could be
included by combining the Super-Kamiokande Electron Scattering measurement with
the SNO Neutral Current measurement. However, since the uncertainty on the NC measurement
dominates this would produce a similar sized band.}.  From this plot, we can
derive the {\em lower bound} on $Y_e\rho~\vert^{^8\text{B}}_{eff}$
 which is 40 g/cm$^3$ for any value of
$\theta_\odot$ at 95 \% CL.  Future reactor neutrino oscillation
experiments~\cite{SADO} can perform a 2-3\% measurement of $\sin^2 \theta_\odot$.
 The 68\% range of
$\sin^2\theta_\odot$ is indicated by
vertical lines in this figure.
However, such precision on $\sin^2 \theta_\odot$ will not reduce the allowed values
for $Y_e\rho~\vert^{^8\text{B}}_{eff}$ unless the error 
on the measured value of CC/NC is reduced.   

A three neutrino analysis is needed if $U_{e3} \neq 0$ and this
 can be performed using  Eq.~(\ref{Aeff-SNO}) with the
following replacements, 
\begin{eqnarray}
\theta_\odot \to \theta_{12}, \quad
\delta m_\odot^2 \to  \delta m_{21}^2/ \cos^2\theta_{13} \quad & \text{and} & \quad
\frac{\text{CC}}{\text{NC}} \to  
\frac{\text{CC}}{\text{NC}} \; \frac{1}{\cos^4\theta_{13}}.
\end{eqnarray}

A weak upper bound could be derived using a precision measurement of the $^7$Be and/or pp
electron neutrino survival probability in a similar fashion.  As $A_{eff}$ gets larger,
 the  fraction of $\nu_2$ gets larger, see Eq.(\ref{otherapprx}), and hence the 
electron neutrino survival probability gets smaller for fixed values of the mixing parameters.  
The upper bound arises when this survival probability is below the measured survival probability
at some confidence level,
assuming that the mixing parameters have been determined independent of these solar neutrinos.

\section{Summary and Conclusions}
We have performed an extensive analysis of the mass eigenstate
fractions of $^8$B solar neutrinos using only two mass eigenstates
($\sin^2 \theta_{13}=0$) and with three mass eigenstates ($\sin^2
\theta_{13} \neq 0$).  In the two neutrino analysis the
$\nu_2$-fraction is 91 $\pm$ 2\%.
The remaining 9 $\mp$ 2\% is, of course, in the $\nu_1$ mass
eigenstate.  With these fractions in hand, which are primarily
determined by the solar $\delta m^2$ measured by the KamLAND
experiment, the sine squared of the solar mixing angle is simply
related to CC/NC ratio measured by the SNO experiment.  For
completeness the mass eigenstate fractions for $^7$Be and pp are also
given.

Allowing for small but non-zero $\sin^2 \theta_{13}$, in a full three
neutrino analysis, we found very little change in the fraction of
$\nu_1$'s.  This implies, since the $\nu_3$ fraction is $\sin^2
\theta_{13}$, that the $\nu_2$ fraction is reduced by $\sin^2
\theta_{13}$.  That is, the $\nu_2$-fraction is
\begin{eqnarray}
91 \pm 2 - 100 \sin^2 \theta_{13} ~\% \quad \text{at the 95\% CL.}
\end{eqnarray}
Since the CHOOZ experiment constrains the value of $\sin^2 \theta_{13}
< 0.04$ at the 90\% CL  this places a lower bound on the $\nu_2$
fraction of $^8$B solar neutrinos in the mid-eighty percent range
making the $^8$B solar neutrinos the purest mass eigenstate neutrino
beam known so far, and it is a $\nu_2$ beam!

As an example of the use of these mass eigenstate fractions, we have
shown that for the $^8$B neutrinos observed by the SNO experiment, the
$U_{e2}$-element of the MNS matrix is given by
\begin{eqnarray}
\vert U_{e2} \vert^2 
 \approx & \sin^2\theta^{^8B}_\odot  + (0.53 ^{+0.06}_{-0.04})\sin^2 \theta_{13}.
\end{eqnarray}
Where $\sin^2\theta^{^8B}_\odot$ is the sine squared of the
solar mixing angle determined by using a two neutrino analysis of the  $^8B$ neutrinos
plus KamLAND. 
An analysis for this $\sin^2\theta^{^8B}_\odot$ obtained from the SK, SNO and KamLAND data
\cite{SKanalysis}
gives $\sin^2\theta^{^8B}_\odot = 0.30^{+0.11}_{-0.08}$ at the 95\% CL.
 With the data currently available this is our best estimate of $\vert
U_{e2} \vert^2$ and is the most accurately known MNS matrix element.

Finally, we have also demonstrated the possibility of probing the solar
interior by $^8$B neutrinos. We have derived a lower bound on the
average electron number density over the region where the solar $^8$B neutrinos are
produced which is 50\% of the Standard Solar model value.

 \vspace{-0.3cm}
 \begin{acknowledgments} 
   \vspace{-0.3cm} This work was supported by Funda\c{c}\~ao de Amparo
   \`a Pesquisa do Estado de S\~ao Paulo (FAPESP) and Conselho Nacional
   de Ci\^encia e Tecnologia (CNPq).  Fermilab is operated under DOE
   contract DE-AC02-76CH03000.  Two of us (H.N. and R.Z.F.) are
   grateful for the hospitality of the Theoretical Physics Group of the Fermi
   National Accelerator Laboratory during numerous visits. 
   R.Z.F. is also grateful to  
   the Abdus Salam International Center for Theoretical Physics where 
   the final part of this work was performed.
   We thank Marc  Pinsonneault, Hisakazu Minakata and Alexei Smirnov for discussions.
   
   \end{acknowledgments}

\vspace{1cm}


\end{document}